\documentclass[twocolumn]{revtex4}
\usepackage[colorlinks,bookmarksopen,bookmarksnumbered,citecolor=magenta,urlcolor=red]{hyperref}
\usepackage{graphicx}
\usepackage{amsmath}
\usepackage{natbib}
\usepackage{subfigure}
\usepackage{xcolor}
\usepackage[percent]{overpic}
\usepackage{tikz}
\usepackage{amsmath}
\usepackage{url}
\usepackage{times}
\usepackage{epstopdf}
\begin{document}
\bibliographystyle{abbrvnat}
	
\title{Eddy induced trapping and homogenization of freshwater in the Bay of Bengal}
	
	%----------------------------------------------------------------------------------------
	%	AUTHORS AND AFFILIATIONS
	%----------------------------------------------------------------------------------------
	
	% Use \author{\altaffilmark{}} and \altaffiltext{}
	
	% \altaffilmark will produce footnote; matching \altaffiltext will appear at bottom of page.

\author{Nihar Paul$^{1}$, Jai Sukhatme$^{1,2}$, Debasis Sengupta$^{1,2}$ and Bishakdatta Gayen$^{1,3}$}
%\author{Nihar Paul\affil{1}, Jai Sukhatme\affil{1,2}, Debasis Sengupta\affil{1,2} and Bishakhdatta Gayen\affil{1,3}}

\affiliation{$^1$Centre for Atmospheric and Oceanic Sciences, Indian Institute of Science, Bangalore - 560012, India.}
\affiliation{$^2$Divecha Centre for Climate Change, Indian Institute of Science, Bangalore - 560012, India.}
\affiliation{$^3$Mechanical Engineering Department, University of Melbourne, Australia.}	
	%----------------------------------------------------------------------------------------
	%	ABSTRACT
	%----------------------------------------------------------------------------------------
	
%	% Do NOT include any \begin...\end commands within the body of the abstract.
%	

\begin{abstract}
\noindent 
Freshwater from rivers influences Indian summer monsoon rainfall and regional tropical cyclones by shallowing the upper layer and warming the subsurface ocean in the Bay of Bengal. Here, we use {\it in situ} and satellite data with reanalysis products to showcase how river water can experience a significant increase in salinity on sub-seasonal timescales. This involves the trapping and homogenization of freshwater by a cyclonic eddy in the Bay. Specifically, in October 2015, river water is shown to enter a particularly long-lived eddy along with its attracting manifolds within a period of two weeks. The eddy itself is quite unique in that it lasted for 16 months in the Bay where average lifespans are of the order of 2-3 months. This low salinity water results in the formation of a highly stratified surface layer. In fact, when freshest, the eddy has the highest sea-level anomalies, spins fastest, and supports strong lateral gradients in salinity. Subsequently, observations reveal progressive homogenization of salinity and relaxation of sea-level anomalies and salinity gradients within a month. In particular, salty water spirals in, and freshwater is pulled out across the eddy boundary. Lagrangian experiments elucidate this process, whereby horizontal chaotic mixing provides a mechanism for the rapid increase in surface salinity on the order of timescale of a month. This pathway is distinct from vertical mixing and likely to be important in the eddy-rich Bay of Bengal.
\end{abstract}
%	
%	% The body of the article must start with a \begin{article} command
%	% \end{article} must follow the references section, before the figures and tables.
%\vskip 0.25 truecm
%
\maketitle
\section{Introduction}
\noindent 
The Bay of Bengal (BoB) is a semi-enclosed basin lying between 6-22$^\circ$N, 80-100$^\circ$E, connected to the equatorial Indian Ocean to the south. The Ganga-Brahmaputra-Meghna (GBM), Irrawaddy, Godavari, and Mahanadi river systems are the major sources of fresh water to the BoB \cite{papa2012ganga}. 
The climatological annual mean discharge from the GBM and Irrawaddy, the largest of these rivers, is approximately $8.7\times10^4$ m$^3$s$^{-1}$ and $3.4\times10^4$ m$^3$s$^{-1}$; about 70\% of discharge comes in the summer monsoon season June-September \cite{dai2002estimates,papa2012ganga,chait}. The freshwater from rivers is stirred into the interior of the Bay of Bengal by large-scale circulation, mesoscale eddies, and directly wind-driven flow \cite{sree2018subseasonal}. In the open ocean, river water forms a shallow, low-salinity layer with strong density stratification at its base \cite{sengupta2016near}. Apart from a well-defined seasonal cycle \cite{rao}, lateral advection gives rise to intraseasonal variability in near-surface salinity, as seen in Argo float and satellite data \cite{Param, Grun,trott2019}.
Freshwater in the Bay of Bengal has profound impacts on regional climate, not only does it affect the local circulation \cite{shankar2002monsoon} and sea surface conditions \cite{raghu,seo}, the shallow stratification and thin mixed layers in turn influence regional cyclones \cite{neetu} and the monsoon itself \cite{shenoi,samanta2018impact}.

\noindent
The seasonally reversing East India Coastal Current (EICC) flows northward along the western boundary of the Bay in spring, and southward in autumn \cite{potemra1991seasonal,schott2001monsoon}. {\it In situ} data, as well as model simulations, show that the EICC transports freshwater from the GBM, Mahanadi, and Godavari rivers in the form of a narrow, fresh ``river in the sea" in the post-summer monsoon season \cite{shetye1996hydrography,vinayachandran2005bifurcation,chait}. The southward transport of freshwater by the EICC has been analyzed in numerical models \cite{vinaykurian,vinayravi}; in particular, recent efforts suggest that vertical diffusion plays an important role in the gradual increase of surface salinity on a seasonal timescale \cite{akhil2014modeling,benshila2014upper}. While dynamical mechanisms are not particularly clear, on shorter timescales, Lagrangian salinity change maps from August to October of 2013 also show an increase in saltiness with southward advection \cite{Amala}. 
In this context, the role of eddies in stirring the salinity field has been recognized \cite{kumar2013thermohaline,fournier2017modulation,sree2018subseasonal}. In fact, eddy induced variability in sea surface salinity has been recently observed in other ocean basins such as the tropical Pacific \cite{delcroix} and the Arabian Sea \cite{trott}.

\noindent
The fact that eddies are likely to play an important role in the evolution of the salinity field is not surprising given their influence on the mixing of passive fields on the surface of the Bay \cite{paul2020seasonality}, as well as the broader observation that they trap and transport salt and heat in the global oceans \cite{dong2014global}. Furthermore, eddies are ubiquitous in the BoB \cite{dandapat2016mesoscale} and their properties have been documented on intraseasonal and interannual time scales  \cite{chen2012features,cheng2013intraseasonal,subrahmanyam2018detection,trott2019detection}. 
In this work, we examine the interaction of freshwater and a cyclonic eddy in the BoB. In particular, our interest is on timescales of the order of a month, and we showcase the trapping and progressive homogenization of low salinity water that enters the Bay in the postmonsoon season. The trapping of freshwater is explained from a dynamical systems perspective by appealing to the eddy's attracting manifolds. Then, the exchange of material by means of extended salty (fresh) filaments being wrapped into (out of) the eddy results in the homogenization of freshwater that is quantified by vertical profiles and horizontal cross-sections in and across the eddy. This chaotic mixing of fresh and salty water is elucidated via Lagrangian passive tracer simulations. Further, these processes are also identified in reanalysis data. Finally, the results are discussed and a hypothesis on the dynamical cause of horizontal mixing is presented.

\section{Data sources and methods}
\noindent
A variety of {\it in situ}, satellite and reanalysis data are used in this study: Daily mean sea level anomalies and surface geostrophic currents  (MSLA-UV; $0.25^\circ\!\times 0.25^\circ$) are from AVISO. Daily sea surface temperature (SST) is from the Group of High Resolution Sea Surface Temperature (GHRSST) product that is produced at $0.25^\circ\!\times0.25^\circ$ resolution, and nighttime SST from Advanced Very High-Resolution Radiometer (AVHRR) instrument of METOP2 satellite on a 1.1 km grid. 8-day running mean sea surface salinity (SSS) estimates are from the Soil Moisture Active Passive (SMAP) satellite \cite{fore2016combined} at 60 km resolution, interpolated to a $0.25^\circ\!\times0.25^\circ$ grid. Total surface currents are from the Bay of Bengal current and advection estimation (BoBcat), based on a combination of AVISO geostrophic currents and directly wind-forced Ekman currents at 1 m depth on a grid resolution of $0.25^\circ\!\times0.25^\circ$ \cite{sree2018subseasonal,buckley2020impact}. Temperature and salinity profiles are from Argo floats with 5-day sampling, from NOAA's Atlantic Oceanographic and Meteorological Laboratory \cite{argo2000argo}. We also use drifter data from the Surface Velocity Program (SVP), NOAA’s Global Drifter Program \cite{centurioni2019global}.

\noindent
The ocean reanalysis is the GLORYS12V1 global product that uses the NEMO ocean model (1/12$^\circ$ horizontal resolution, order 1 m vertical resolution in the upper ocean, 50 vertical levels) surface forcing from ECMWF ERA-Interim reanalysis, covering the satellite altimetry era 1993-2018. Observations are assimilated using a reduced-order Kalman filter, with a 3D-VAR scheme used to correct for the slowly-evolving large-scale biases in temperature and salinity \cite{lellouche2018recent,chassignet2018new}.

\noindent
The Lagrangian advection of tracers is performed by BoBcat surface currents from a given initial position by integrating the surface flow field using a Runge-Kutta fourth-order method. The velocities are interpolated using a bi-linear interpolation scheme to a grid of resolution 0.01$^\circ\times0.01^\circ$. The equations involved are described in the supplementary material text S1 (Equation 1). Lagrangian measures such as attracting Lagrangian Coherent Structures (a-LCSs) or backward Finite Time Lyapunov Exponents (b-FTLEs) \cite{wiggins2005dynamical,haller2011lagrangian,onu2015lcs} have also been computed to understand the pathway of freshwater trapping further. Details of the calculation of these metrics are provided in the supplementary material (text S1). Given the timescales of interest here, the backward advection time has been taken as 20 days to calculate b-FTLEs (a-LCSs).

\section{The eddy and the entry of freshwater}
\begin{figure*}
	\centering
	\includegraphics[scale=1]{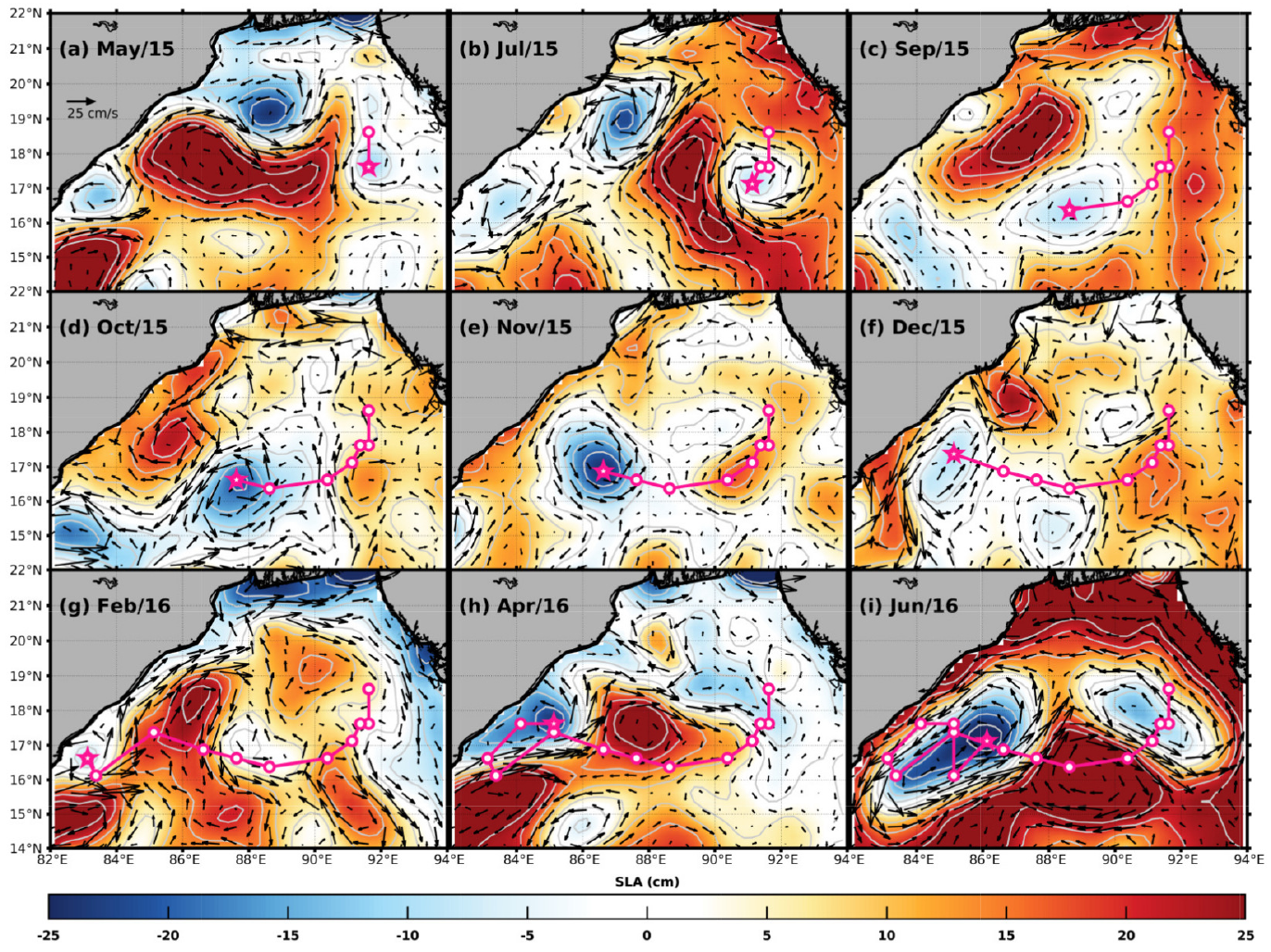}
	\caption{(a)-(i) Mean sea level anomaly (SLA) contours in units of centimeters with geostrophic velocity quivers overlaid on 1st day of the month for May, July, September, October, November, December 2015, and February, April, June 2016. The contours of SLA are in the range of -25 cm to 25 cm with 5 cm intervals. The track of the cyclonic eddy is shown with ``star'' indicating the SLA minimum of the eddy and ``dot" denoting the position in preceding months from its origin.}
	\label{fig:a1}
\end{figure*}

\noindent
Snapshots of the particular eddy for every month that we use to highlight interaction with freshwater is shown in Figure \ref{fig:a1}. In particular, we show the eddy center (star), sea-level anomaly, and geostrophic currents. Consistent with prior work \cite{cheng2018dynamics}, this eddy was born at the eastern edge of the Bay. After its genesis (around April 2015), the eddy moved southwest towards the central BoB, then took a northwestward route from September 2015 onward (Figure \ref{fig:a1}(a),(b),(c)). After the monsoon, the eddy remained close to the western boundary of BoB. Interestingly, the eddy appeared to decay in size towards the end of 2015 but re-intensified, and its size increased on the arrival of the northward-flowing EICC \cite{gangopadhyay2013nature} during spring of 2016 (Figure \ref{fig:a1}(g)). Finally, the eddy moved eastward, off of the western boundary, towards the central BoB and dissociated around June 2016 (Figure \ref{fig:a1}(h),(i)). The westward (eastward) drift of the eddy early (late) in its life is succinctly captured by the Hovm\"oller diagram in Figure S1(a). In particular, the speed of translation in these two periods is seen to be approximate $-6$ cm/s and $4$ cm/s, respectively. Further, while near the western coast, the Rossby number ($Ro$) --- defined as $\xi/f$, where $\xi$ is the relative vorticity of the eddy and $f$ is the Coriolis parameter --- of this eddy was approximately $0.2$ suggestive of geostrophic balance. Given that the average life span of eddies in the BoB is about two-three months \cite{chen2012features,cheng2018dynamics}, this particular eddy was remarkable in that it lasted for about sixteen months. Indeed, such a long life span of eddy can be sustained in BoB has not been highlighted so far in the literature as far as we are aware. 

\begin{figure*}
	\centering
	\includegraphics[scale=0.5]{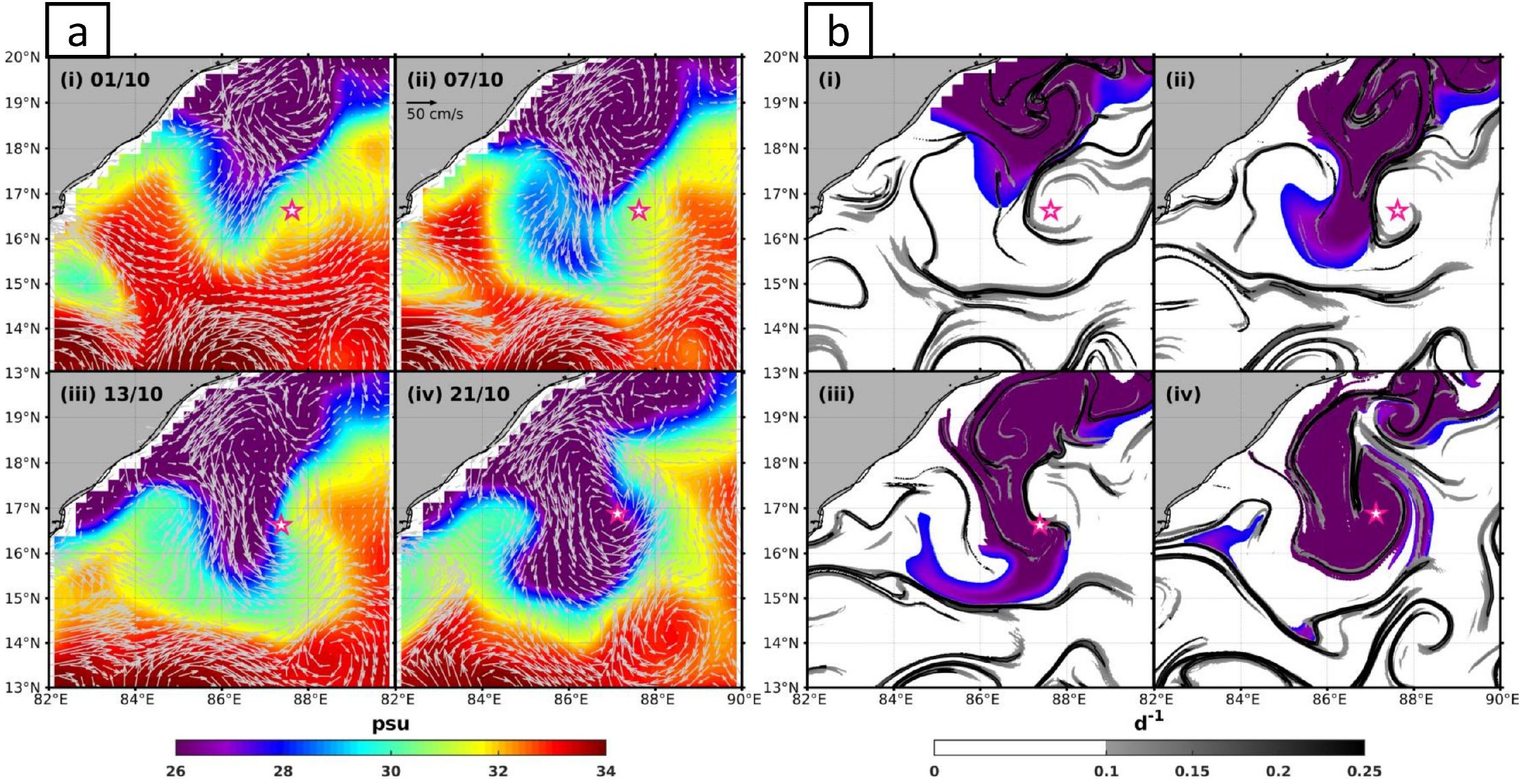}
	\caption{(a) Sea Surface Salinity (SSS) with BoBcat current quivers on 01/10, 07/10, 13/10, 21/10, respectively. (b) Advected passive scalar maps with tracer initialized to SSS$<28$ psu on 01/10 for days as in (a). In addition, attracting Lagrangian Coherent Structures (a-LCS) or b-FTLE computed by integrating backward for 20 days are shown on the top of the tracer field. Only values above $0.1$ day$^{-1}$ (stronger stable manifolds) are shown.}
	\label{fig:a2}
\end{figure*}

\noindent
In the post-monsoon season, the eddy translated along western boundary of the Bay and trapped freshwater as shown in the SSS maps with surface currents in Figure \ref{fig:a2}(a). Starting in early October, freshwater is pulled inward from the western boundary and enters the eddy over a period of about 10 days. 
Coincidentally, five drifters were also trapped and formed loops within the eddy during this period as shown in Figure S1(b). Signs of chaotic advection \cite{ottino1989kinematics} of the salinity field, that are visible via the stretching of the salinity field in Figure \ref{fig:a2}(a), are brought out in Figure \ref{fig:a2}(b) via a tracer experiment \cite{paul2020seasonality}, wherein a passive field is initialized with the same value as the freshwater salinity (SSS$\le 28$ psu) on 01/10. The backward finite time Lyapunov exponents (b-FTLEs; see Supplementary Material text S1 for definition and computation details) \cite{perez2014mixing,mathur2019thermal} for the BoBcat surface flow are also shown for clarity. As is evident, the freshwater (and tracer) is pulled in along these attracting manifolds \cite{d2004mixing,lehahn2007stirring} and ``fills" the eddy in a span of about 10 days. Thus, in less than two weeks, the eddy which contained salty water (around 01/10), pulls in freshwater that forms a layer on its surface (by 21/10). The role of mesoscale features in restricting freshwater to the north Bay has been suggested in other tracer experiments too \cite{benshila2014upper}.

\begin{figure*}
	\centering
	\includegraphics[scale=0.85]{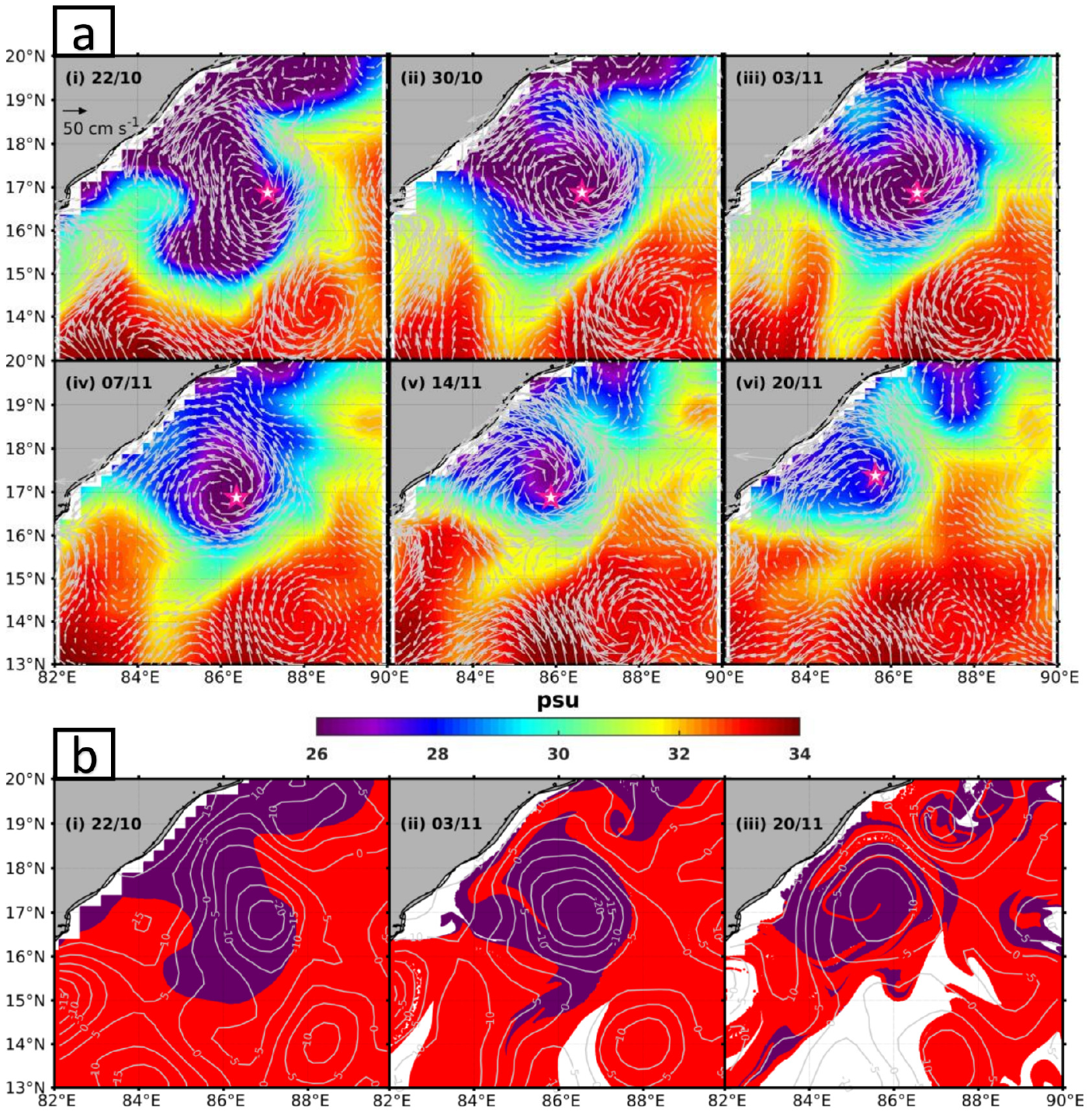}
	\caption{(a)(i)-(vi) shows Sea Surface Salinity (SSS) with BoBcat current quivers on 22/10, 30/10, 03/11, 07/11, 14/11, 20/11, respectively. Star marks the center of the eddy. (b) Tracers with SSS $\le$ 28 ($>$28) psu initialized on 22/10/2015 are marked in violet (red) colors. These are advected forward in time by BoBcat currents and shown for 22/10, 03/11, and 20/11 of 2015 with contours of SLA overlaid on the top.}
	\label{fig:a3}
\end{figure*}

\begin{figure*}
	\centering
	\includegraphics[scale=0.45]{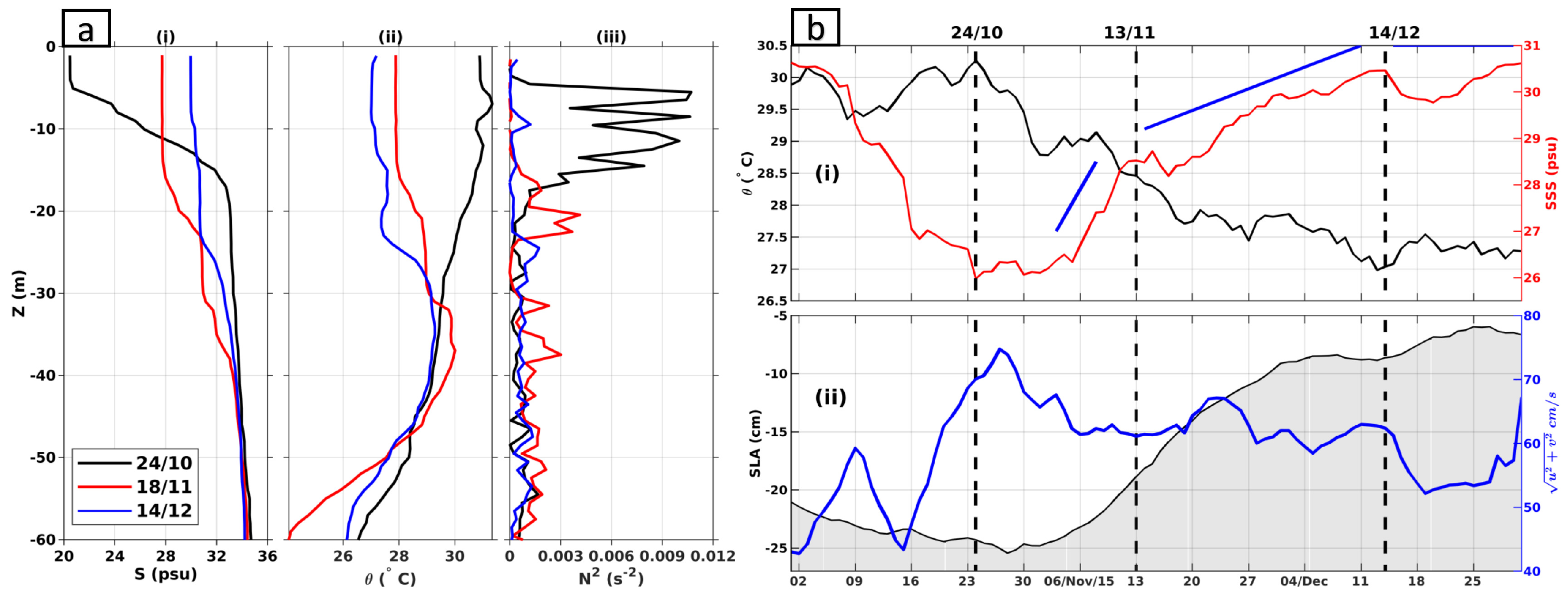}
	\caption{(a)(i),(ii),(iii) shows S (psu), $\theta$ ($^\circ$ C) and N$^2$ (s$^{-2}$) with depth for upper 60 m on 24/10, 18/11 and 14/12 from Argo AOML-5904302, respectively. (b) Daily time series of (i) mean $\theta$, SSS (with straight lines to guide the eye) and (ii) maximum current speed and minima of SLA over a circle of diameter 300 km around the eddy center from 01/10/2015 to 31/12/2015.}
	\label{fig:a4}
\end{figure*}

\begin{figure*}
	\centering
	\includegraphics[scale=0.8]{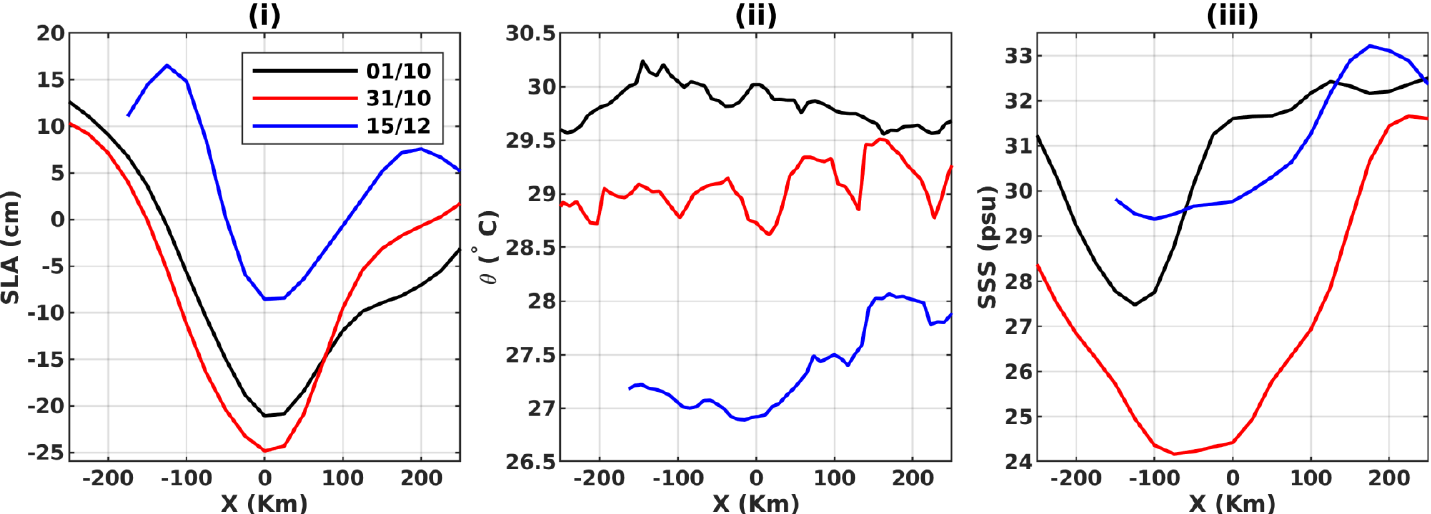}
	\caption{(i), (ii), (iii) shows SLA, $\theta$ and SSS cross-sections through the center of the eddy on 01/10, 31/10, and 15/12 representing pre-freshening, freshening, and post-freshening days, respectively.}
	\label{fig:a5}
\end{figure*}

\section{Homogenization of freshwater in the eddy}

\noindent
The freshwater that entered the eddy by means of horizontal chaotic advection remains trapped for more than a month as seen in the SSS maps in Figure \ref{fig:a3}(a). Figure \ref{fig:a3}(b) shows snapshots of the evolution of a passive field (SSS$\le$28 psu; violet color) within the eddy for this time period. As with SSS, the passive tracer remains contained in the eddy for the most part, though there is a systematic exchange of material across the eddy boundary. Specifically, some salty water (SSS$>$28 psu; marked by red color) spirals into the eddy while freshwater (violet) is stretched out in the form of extended filaments that mix into the exterior Bay. For example, the intrusion of red filaments that get wrapped into the eddy and violet streaks that are stretched out of the eddy can clearly be seen in Figure \ref{fig:a3}(b) on 03/11 and 20/11. This exchange of material across the ``kinematic" boundary of the eddy is a hallmark of chaotic mixing (see, for example, \citep{JL}, for a discussion of this mixing process and an example from the atmosphere) and is anticipated to homogenize salinity inside the eddy. In fact, as can be seen from the SSS values in Figure \ref{fig:a3}(a), the freshwater on 22/10 is much saltier after a month's time. 

\noindent
The vertical structure of water mass is obtained from Argo AOML-5904302 data which was trapped in this eddy from 24/10 to 24/12 (Figure S1(b)). In particular, the evolution of potential temperature ($\theta$), salinity (S), potential density ($\sigma_\theta$), square of the Brunt-V\"{a}is\"{a}l\"{a} frequency (N$^2$ $=-\frac{g}{\bar{\rho}}\frac{d\rho}{dz}$) where $\theta$, $\sigma_\theta$ and N$^2$ has been computed using Gibbs-SeaWater Oceanographic Toolbox \cite{mcdougall2011getting}, and is shown in Figure S2. The homogenization suggested by Figure \ref{fig:a3}(a) is confirmed via this {\it in situ} data by comparing vertical profiles of $\theta$, S and N$^2$ in the early (24/10), middle (18/11) and late (14/12) parts of the aforementioned period. Specifically, as per Figure \ref{fig:a4}(a), on the entry of freshwater (24/10), we clearly see a very shallow layer (about 5 m) of low salinity water, a small temperature inversion, and high buoyancy frequency below the base of this fresh layer. This results in a stratified upper layer in the eddy with N$^2_{max}=1.07\times 10^{-2}$s$^{-2}$. In a few weeks, by 18/11, the surface salinity (temperature) shows signs of increasing (decreasing) and the buoyancy frequency has reduced (N$^2_{max}=4.1\times 10^{-3}$s$^{-2}$). In about a month's time, i.e., on 14/12, it becomes clear that surface salinity has increased by almost 10 psu in this point measurement and its vertical structure is weaker. Further, the surface of the eddy has cooled and there is a marked inversion in temperature at about 25 m \cite{thad}. The buoyancy frequency has also reduced significantly (approximately N$^2_{max}=1.8\times10^{-3} $s$^{-2}$) and is fairly uniform with depth. 
Thus, these profiles of water mass clearly indicate a progressive homogenization of fields and a decrease in vertical gradients over the course of a month.

\noindent
The evolution of mean surface potential temperature ($\theta$), SSS, sea-level anomaly of the eddy center, and a maximum speed of the surface currents over a circle of diameter $300$ km across the eddy center is shown in Figure \ref{fig:a3}b(i),(ii), respectively as it moves along the western coast of the BoB. Here, we see that during the entry of freshwater (09/10 till 30/10), SSS drops by 5 approximately psu. Then, consistent with the Argo float data, surface salinity starts increasing. The initial rate of increase is relatively high rate till about 13/11 (approximately 3 psu in a week), thereafter, the rate of increase of salinity slows down, and by 14/12 the mean eddy salinity is about 30-31 psu. This indicates that the homogenization of salinity possibly involves multiple stages with different mixing rates. Further, along with this increase in SSS over a month's time, there is systematic surface cooling from about $30.5$ to $27.5^\circ$C and an increase in the sea-level anomaly of the eddy center within the eddy. 

\noindent
Longitudinal variations across the center of the eddy of sea level anomaly, $\theta$, and SSS and during pre-freshening (01/10), freshening (31/10), and post freshening days (15/12) are shown in Figure \ref{fig:a5}. The sea-level anomaly is minimum (about $-25$ cm) and the rate of spinning is at its fastest when freshest water is in the eddy as compared to pre and post freshening periods. Consistent with vertical profiles in Figure \ref{fig:a5}a the potential temperature decreases across the eddy with time after the entry of freshwater. Interestingly, as seen in Figure \ref{fig:a5}, there is a strong gradient in salinity across the eddy of approximately 6 psu in 200 km on 31/10 when freshwater filled the eddy. This horizontal gradient relaxes in a month (by 15/12), and not only is the salinity is much higher, but sea-level anomalies also become smaller (about $-10$ cm) and the rate of spinning slows down. Thus, the surface properties and horizontal sections go hand in hand with Argo data and support the progressive homogenization of freshwater in the eddy. 

\section{Ocean reanalysis and discussion}

\begin{figure*}
	\centering
	\includegraphics[scale=1]{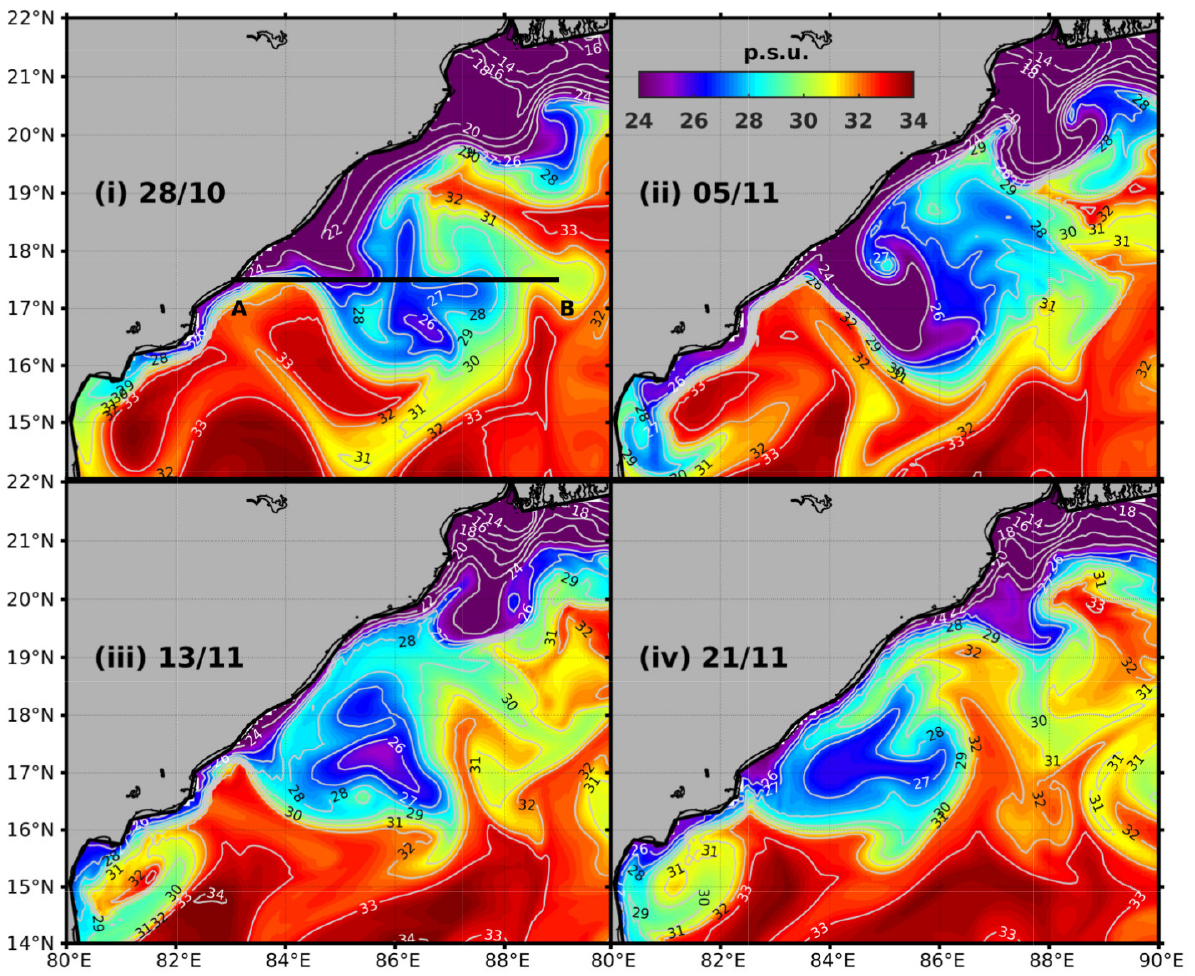}
	\caption{(i)-(iv) SSS (psu) (with contours) at 0.5 m from NEMO reanalysis from 28/10 to 21/11 (freshwater trapping days) with 8-day interval. AB denotes the cross-section of the eddy from 83$^\circ$E-89$^\circ$E shown in subpanel (i).}
	\label{fig:a6}
\end{figure*}

\begin{figure*}
	\centering
	\includegraphics[scale=0.48]{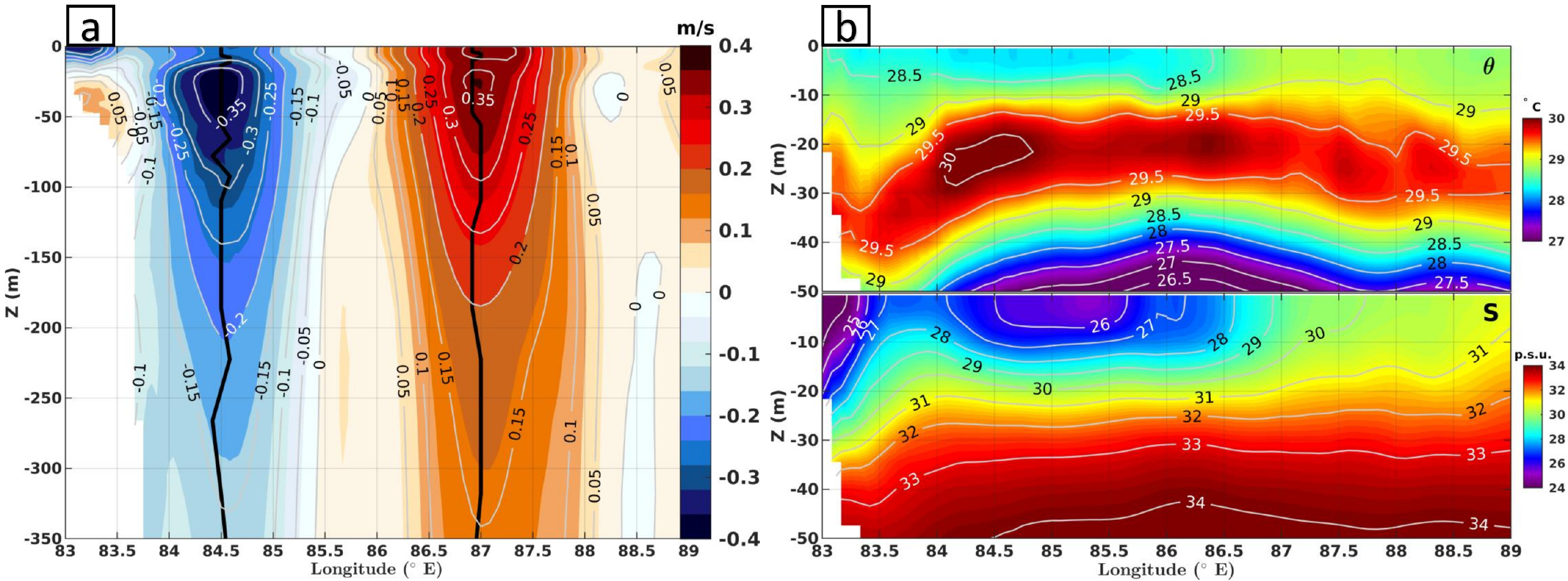}
	\caption{(a) Mean meridional velocity profile along the longitudinal section AB of the eddy up to a depth of 350 m shown in Figure \ref{fig:a6}(i) averaged over freshwater trapping days (28/10--21/11) from NEMO reanalysis data. The vertical black lines indicate the contours of maximum speed on each side of the lobes of the eddy. (b) Variation of mean potential temperature ($\theta$) (upper panel) and mean salinity (lower panel) within upper 50 m of the eddy averaged for the same section and time period as in (a).}
	\label{fig:a7}
\end{figure*}

\noindent
In the Sections so far, we saw the trapping and homogenization of freshwater via {\it in situ} and satellite data. Here, we check if similar features can be detected in the NEMO reanalysis product. Quite clearly, freshwater can be seen inside the eddy over a similar time period (Figure \ref{fig:a4}a). Further, early on, around 28/10 the water in the eddy is freshest and its saltiness progressively increases over the next three weeks (via the progressively lighter blue colors in Figure \ref{fig:a6}). Given the relatively higher spatial resolution of NEMO reanalysis, we can see the exchange of water mass across the kinematic eddy boundary. This is especially clear on 13/11 and 21/11 where see freshwater (blue) being pulled out in the south-east corner of the eddy while salty water (red) pushes in on its north-east edge. The three-dimensional nature of reanalysis also allows for a clearer view of the eddy itself --- Figure \ref{fig:a7}(a) shows its vertical structure via the meridional velocity (averaged from 28/10 to 21/11) up to a depth of 350 m across the section AB marked in Figure \ref{fig:a6}(i). The jets supported by this eddy have the highest speeds that are approximately 0.35 m/s at a depth of about 20 m, and most of the eddy signature dies out by 300 m. Through this period of trapping, the reanalysis salinity also shows a freshwater layer restricted to within 10 m from the surface (Figure \ref{fig:a7}(b); second panel). Further, the inversion noted due to surface cooling in the {\it in situ} data (Figure \ref{fig:a4}(a)) is also captured here at a depth of approximately 25 m as seen in the first panel of Figure \ref{fig:a7}(b). Thus, in addition to a detailed view of the eddy, while the actual numbers differ slightly, the overall picture of the trapping of freshwater and its progressive homogenization due to lateral mixing is seen in the reanalysis data too.  

\begin{figure*}
	\centering
	\includegraphics[scale=0.8]{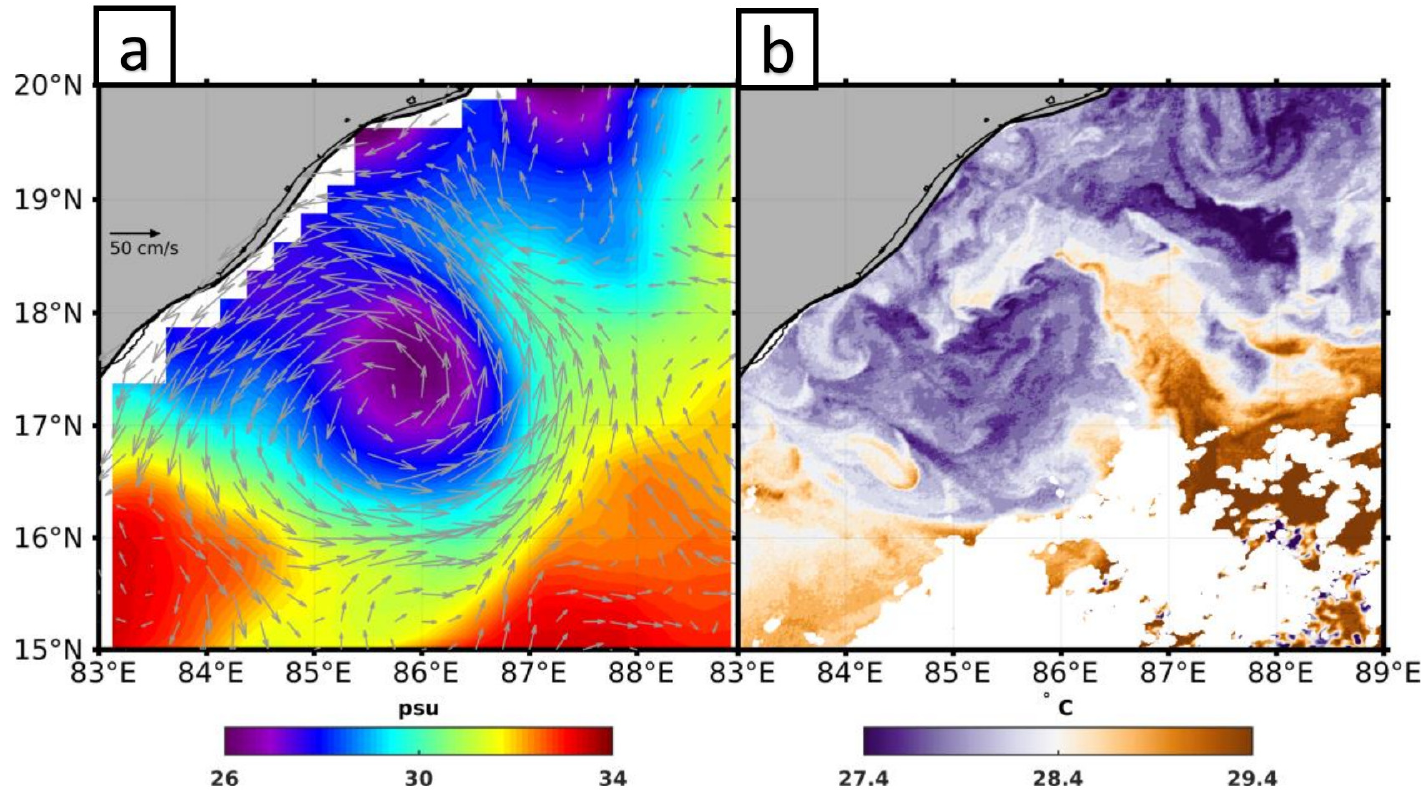}
	\caption{(a) SSS (psu) with BoBcat currents (b) METOP2-AVHRR SST ($^\circ$C) on 13/11/2015.}
	\label{fig:a8}
\end{figure*}

\noindent
Given the inflow of low salinity water to the Bay every year and the ubiquitous presence of eddies on the western coast of the Bay, we suspect that this process of trapping and subsequent homogenization of surface salinity within an eddy on the timescale of a month may not be uncommon. The rate of homogenization is about 2-3 psu in a week which is comparable to mixing during strong wind events, for example, tropical cyclone Phailin in 2013 off the west coast of north BoB caused a change of about 1-4 psu via wind-induced vertical mixing \cite{chaudhuri2019response}. In fact, the increase in SSS by about 5-7 psu in the eddy is of the same order as that noted through the course of the winter season in the northern BoB \cite{akhil2014modeling}. Of course, the outstanding issue that remains is the dynamical cause of mixing in the eddy. Given the strong stratification on the arrival of fresh water, vertical overturning is likely inhibited \cite{thakur2019seasonality}, even with the observed surface cooling \cite{Amala}. 
While the scale of the eddy (approx 300 km) is in a balanced regime \cite{sukhatme2020near}, there are clear indications toward the formation of smaller scales from the SSS and SST fields shown in Figure \ref{fig:a8}(a) and (b). For SSS (Figure \ref{fig:a8}(a)), which is relatively coarse in resolution, salty water can be seen pushing into the eddy on its north-east corner. This is seen in much greater detail with finer scales along with the cold and warm tongues in the higher resolution SST map on the same day in Figure \ref{fig:a8}(b). Indeed, 
lobe like structures can be seen around the eddy on multiple days during the homogenization period (Figure \ref{fig:a3}(a), especially, 30/10, 03/11, and 07/11). 
This is reminiscent of the development of non-axisymmetric features via baroclinic and barotropic instability during the adjustment of rotating-stratified shallow vortices that are of a lower density \cite{saunders1973instability,griffiths-1981,chia1982laboratory,verzicco1997dynamics} or fresher \cite{tartinville1998coastal} than the ambient fluid. 
These features, especially smaller-scale vortices within the parent eddy, are also visible in reanalysis data shown in Figure \ref{fig:a6} and Figure S3. 

\noindent
Indeed, the large gradients observed on the arrival of freshwater are probably even sharper as per ship-based measurements in the BoB \cite{ramachandran2018submesoscale}. At these fine scales of O(1-10 km) frictional as well as diabatic surface fluxes can play an important role in potential vorticity extraction or injection to generate instability and subsequent sub-mesoscale turbulence \cite{thomas2005destruction,d2011enhanced,sarkar2016interplay,thomas2016symmetric,ramachandran2018submesoscale}. In effect, we suspect that the SST in Figure \ref{fig:a8}(b) just provides a glimpse into the horizontal mixing and turbulence triggered by the adjustment process in such freshwater eddies and motivates further study based on {\it situ} observations along with high resolution modeling to understand the mechanism behind the homogenization of salinity at sub-mesoscales. 

\section{Conclusions}
\noindent
By using satellite, {\it in situ} and reanalysis data, we have shown that freshwater discharged into the Bay of Bengal was trapped in a cyclonic mesoscale eddy during the postmonsoon season of 2015. The eddy responsible for trapping freshwater was itself unique in that it lasted for almost sixteen months in the Bay, where the usual lifetimes of eddies are of the order of two months. During its lifetime, the eddy remained close to the Bay's western coast for almost three months in the postmonsoon season. In this period, entry of low salinity water into the eddy began in early October. In particular, freshwater was directed along with the eddy's attracting manifolds and bore the hallmarks of chaotic advection. In a time span of approximately ten days, the eddy, which contained warm and salty water, had a surface layer of cool freshwater. This led to a strongly stratified surface layer, deep sea-level anomalies, high spin rates, and strong lateral gradients in salinity and density across the eddy.

\noindent
The trapped freshwater in the eddy was then observed to be progressively homogenized with its environment. Specifically, satellite maps showed an increase in salinity in about a month after the freshwater entered the eddy. This was corroborated by {\it in situ} vertical profiles from Argo data that showed an increase in surface salinity along with surface cooling and the formation of an inversion layer. Concomitantly, horizontal gradients of salinity and density relaxed over the next month. The homogenization process was illustrated via Lagrangian tracer experiments that showed the horizontal mixing that took place, wherein salty water was advected into the eddy. Freshwater was expelled in thin meandering filaments into the open Bay. The trapping, freshwater surface layer, and its homogenization were also seen in reanalysis data. Thus, %much like recent recognition in the tropical Pacific \cite{delcroix}, 
in contrast to seasonal timescales where vertical diffusion is important,
we have shown that freshwater becomes significantly saltier within a month's timescale, apparently by horizontal mixing when trapped in a mesoscale eddy.
		
\section{Acknowledgement}
We thank Dr. Jared Buckley and Prof. Amit Tandon for sharing the BoBcat dataset for 2015 (\url{https://github.com/jbuckley-BoBcat/BoBcat}). NP would like to acknowledge Dr. J Sree Lekha and Dr. Dipanjan Chaudhuri for discussions. The authors would like to acknowledge the following data sources: National Oceanic and Atmospheric Administration's Atlantic Oceanographic and Meteorological Laboratory, Physical Oceanography Division (\url{https://www.aoml.noaa.gov/phod/gdp/}) for distributing Argo and Drifter data; Copernicus Marine and Environment Monitoring Service (CMEMS) for Ssalto Duacs/gridded multimission altimeter products from AVISO and  GLORYS12V1 product driven by the NEMO model (\url{https://resources.marine.copernicus.eu/}); Physical Oceanography Active Archive Centre for distributing SMAP (Soil Moisture Active Passive) satellite sea surface salinity data (\url{https://doi.org/10.5067/SMP43-3TPCS}) and Group for High Resolution Sea Surface Temperature (GHRSST) data (\url{https://podaac.jpl.nasa.gov/GHRSST/}). We also thank the Divecha Centre for Climate Change, IISc, Bangalore, for providing financial assistance. JS would like to acknowledge support from the University Grants Commission (UGC) for funding via 6-3/2018 under the 4th cycle of the Indo-Israel joint research program. DS acknowledges support from the National Monsoon Mission, IITM, Pune. NP and JS would like to thank Pattabhi Rama Rao E, Dr. N. Srinivasa Rao and Geetha Gujjari of ESSO - Indian National Centre for Ocean Information Services (\url{https://incois.gov.in/}), Ministry of Earth Sciences, Government of India for sharing the METOP2-AVHRR SST data.

\section{Supporting Information for ``Eddy induced trapping and homogenization of freshwater in the Bay of Bengal"}
\subsection{SI: Text}
\noindent 
\textbf{Attracting Lagrangian Coherent Structures (a-LCSs or Backward Finite Time Lyapunov Exponents (b-FTLEs)}: The mixing of freshwater is characterized from a Lagrangian perspective via so-called backward Finite Time Lyapunov Exponents (b-FTLEs) \cite{wiggins2005dynamical}. These are ridges that represent attracting Lagrangian coherent structures in a flow \cite{haller2002lagrangian}. To compute the b-FTLEs, we first advect fluid parcel by integrating the following equations backward in time,

\begin{align}
\frac{d\phi}{dt}=\frac{u(\phi,\lambda,t)}{R\cos(\lambda)},~
\frac{d\lambda}{dt}=\frac{v(\phi,\lambda,t)}{R}.
\label{eqn:1}
\end{align}
Here, $\phi$, $\lambda$, $u$ and $v$ are the latitude, longitude, zonal and meridional velocity, respectively. $R$ is the radius of the earth.
The time span is $t = t_0$ to $t = t_o-\tau$ and the numerical method employed is the 4th order Runge-Kutta scheme. The velocity data $(u, v)$ is given on a fixed grid and the flow has been interpolated by a bilinear interpolation scheme. We then compute the right Cauchy-Green Lagrange tensor $C^t_{t_0}$ associated with the flow map $F^t_{t_0}(\mathbf{x_0})$, which is defined as,

\begin{align}
C^{t}_{t_0}(\boldsymbol{x_0}) = {(\nabla F^t_{t_0}(\boldsymbol{x_0}))}^T \nabla {F^t_{t_0}(\boldsymbol{x_0})}.
\label{eqn:2}
\end{align}
$F^t_{t_0}(\boldsymbol{x_0})$ denotes the position of a parcel at time $t$ backward in time, advected by the flow from  an initial time and position ($t_0,\boldsymbol{x_0}$). $C^{t}_{t_0}(\boldsymbol{x_0})$ is symmetric and positive definite, its eigenvalues ($\lambda's$) and eigenvectors ($\xi's$) can be written as,

\begin{align}
C^{t}_{t_0}(\boldsymbol{x_0})=\lambda_i\xi_i,\: 0<\lambda_1\le\lambda_2, i= 1, 2.
\label{eqn:3}
\end{align}  
The gradient of the flow map $\nabla F^t_{t_0}(\boldsymbol{x_0})$ is computed using an auxiliary grid about the reference point \cite{onu2015lcs}, and can be written as,

\begin{align}
\nabla F^t_{t_0}(\boldsymbol{x_0})\approx \begin{pmatrix} 
\alpha_{11} & \alpha_{12} \\
\alpha_{21}& \alpha_{22}
\end{pmatrix},
\label{eqn:4}
\end{align}
where, 

\begin{align}
\alpha_{i,j}\equiv \frac{x_{i}(t;t_0,x_0+\delta x_j)-x_{i}(t;t_0,x_0-\delta x_j)}{2|\delta x_j|}.
\label{eqn:5}
\end{align}
Finally, the largest b-FTLE \cite{haller2002lagrangian,haller2011lagrangian,mathur2019thermal} associated with the trajectory $\boldsymbol{x}(t,t_0,\boldsymbol{x_0})$ over the time interval [$t_0$, $t$] is defined as,

\begin{align}
\lambda_{\tau}(\boldsymbol{x_0})=-\frac{1}{|t-t_0|}\log(\sqrt{\lambda_{\max}[C^{t}_{t_0}(\boldsymbol{x_0})]}).
\label{eqn:6}
\end{align}
The backward integration time $|\tau|=|t-t_0|$ has been taken as 20 days and computed on a finer grid resolution $0.01^\circ\times0.01^\circ$. 

\subsection{SI: Figure (S1, S2, S3)}
\renewcommand{\thefigure}{S1}
\begin{figure*}[ht]

%	\setfigurenum{S1} %%You can change number for each figure if you want, not required. "S" prepended automatically.
	\includegraphics[scale=0.8]{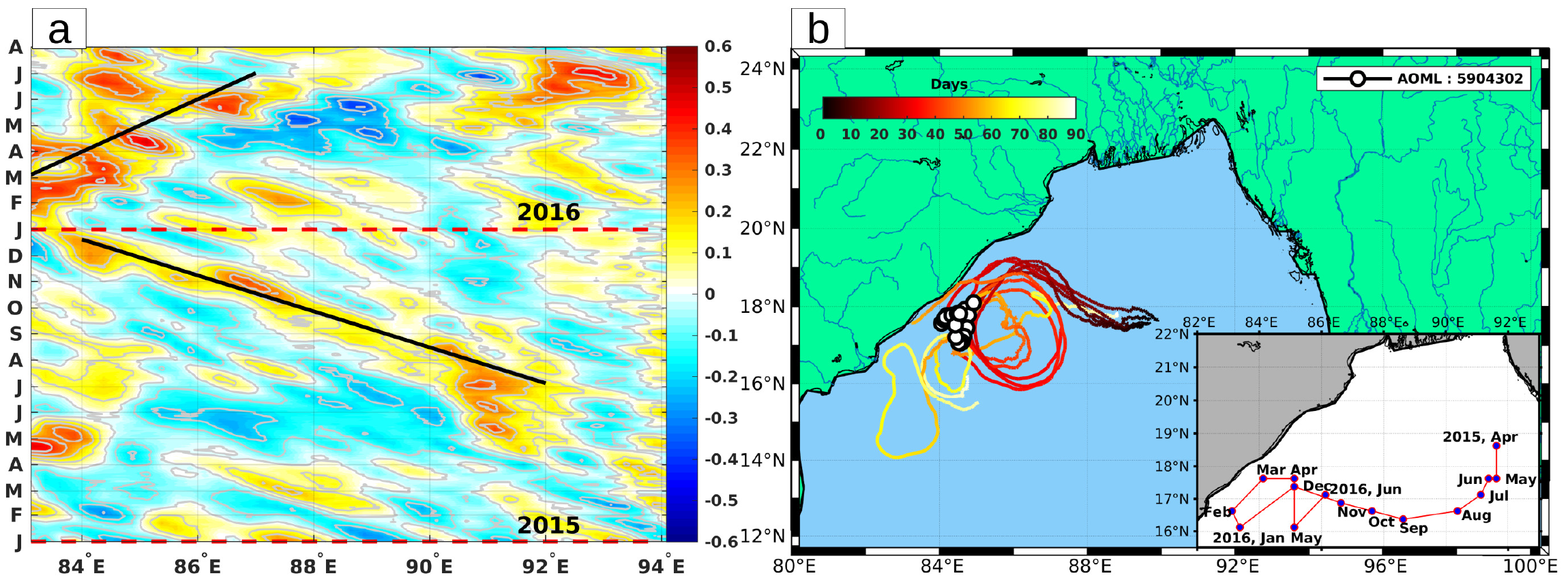}
	\caption{(a) Hovm\"oller diagram of Rossby number ($\xi/f$) computed from geostrophic currents averaged over 16.625$^\circ$N-17.625$^\circ$N shown for the year 2015-2016. (b) Track of Argo float (AOML-5904302) and trajectories of Surface Velocity Program (SVP) ``drifters'' (drouged at 15 m depth) within the eddy from 01/10 to 31/12 of 2015 and entire track of eddy in inset from the first day of April 2015 to June 2016 in an interval of a month.}
	\label{fig1}
\end{figure*}

% \begin{figure}[ht]
% 	\setfigurenum{S3} %%You can change number for each figure if you want, not required. "S" prepended automatically.
% 	\centering
% 	\includegraphics[scale=0.9]{FIG_S3_DRAFT-1.pdf}
% 	\caption{(a) shows SSS (with contours) at 0.5 m from NEMO reanalysis from 28/10 to 21/11 (freshwater trapping days) with 8 day interval. (b) shows the mean meridional velocity profile along the longitudinal section AB of the eddy shown in (a) averaged over freshwater trapping days. It is shown to a depth of 350 m from the surface. The vertical black lines indicates the maximum speed on each lobes of the eddy. (c) shows the variation of mean potential temperature ($\theta$) and mean salinity (S) with upper 50 m of the eddy averaged over the period (freshwater trapping days) along the section AB as in (a).}
% 	\label{fig3}
% \end{figure}

\renewcommand{\thefigure}{S2}
\begin{figure*}[ht]
%	\setfigurenum{S2} %%You can change number for each figure if you want, not required. "S" prepended automatically.
	\includegraphics[scale=0.8]{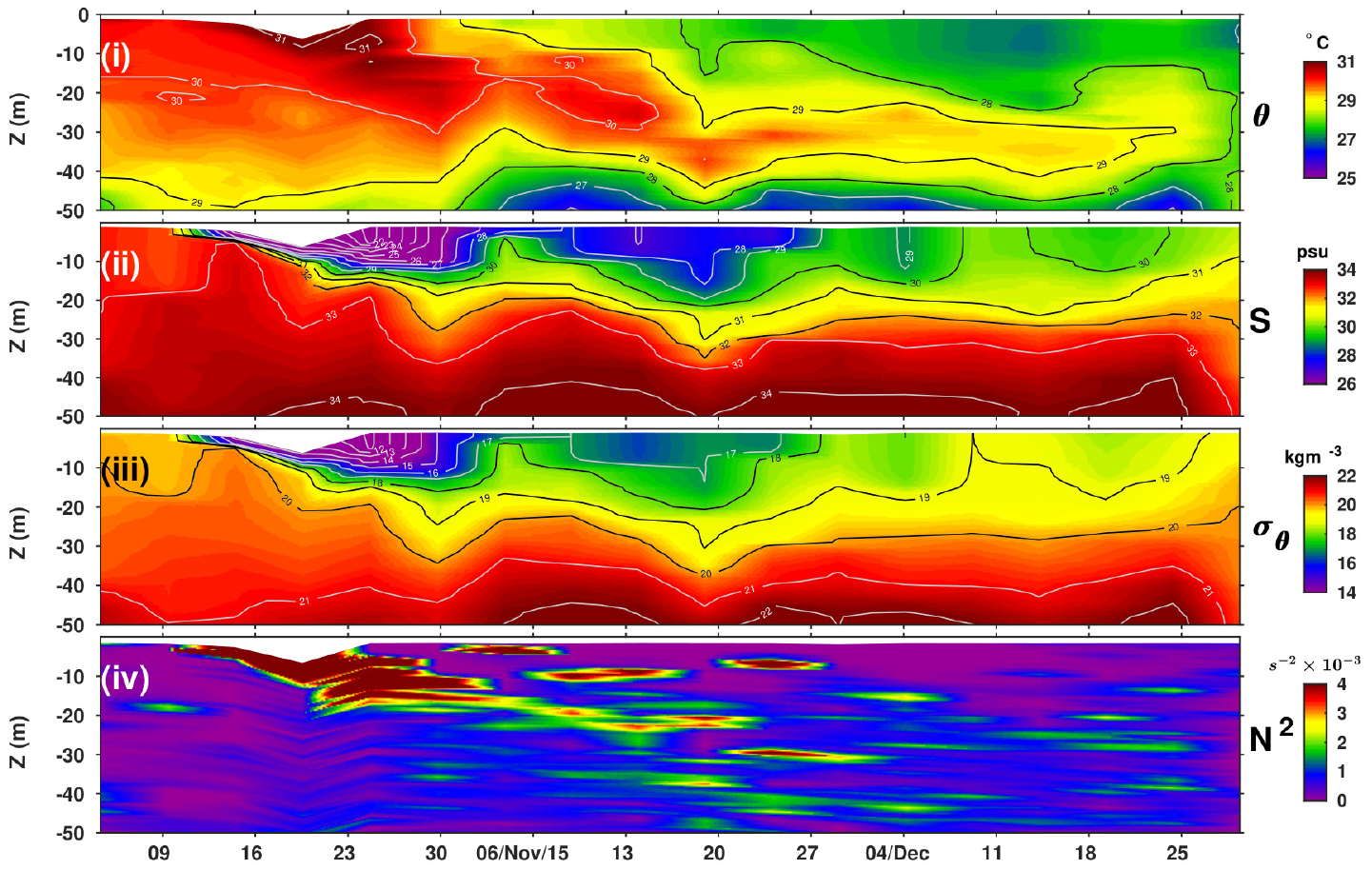}
	\caption{(i) Potential temperature ($\theta$), (ii) salinity (S), (iii) potential density ($\sigma_{\theta}$) and (iv) squared Brunt-v\"{a}is\"{a}l\"{a} frequency (N$^2$), respectively with depth using Argo AOML-5904302 data for the upper 50 m from October to December, 2015.}
	\label{fig2}
\end{figure*}

\renewcommand{\thefigure}{S3}
\begin{figure*}[ht]
%	\setfigurenum{S3} %%You can change number for each figure if you want, not required. "S" prepended automatically.
	\includegraphics[scale=1]{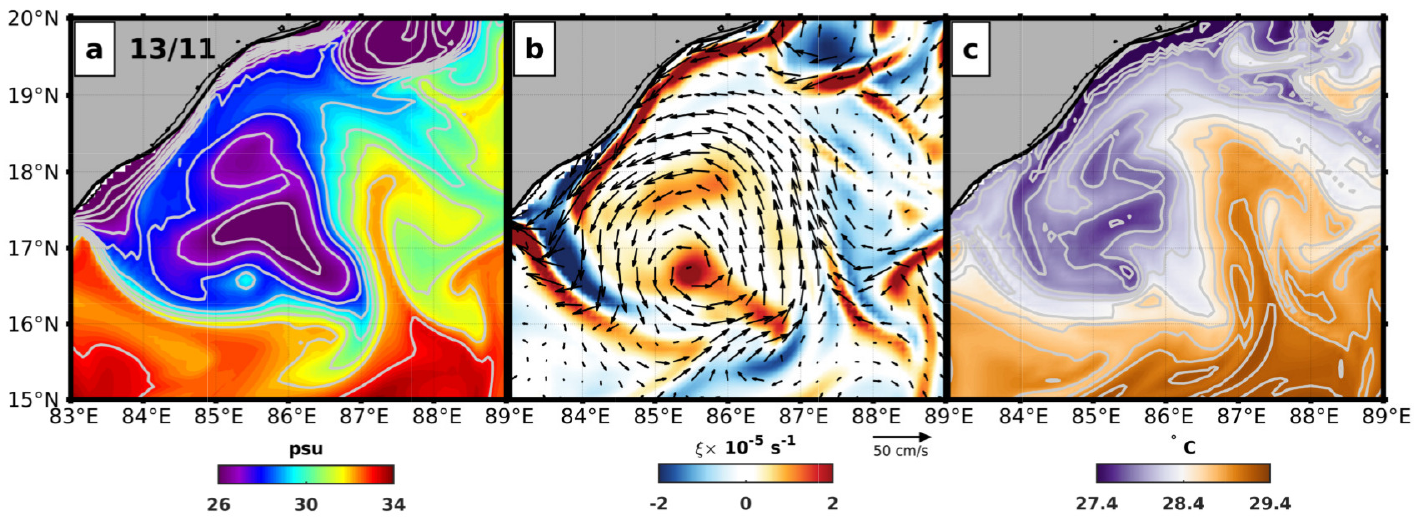}
	\caption{(a) SSS with contours with an interval of 1 psu (b) vorticity with surface current quiver (c) SST with contours with an interval of 0.2$^\circ$C at 0.5 m depth on 13/11/2015 from NEMO reanalysis data.}
	\label{fig3}
\end{figure*}
		
\bibliography{reference}

\end{document}